\documentclass[aps,pre,twocolumn,groupedaddress,showpacs,superscriptaddress]{revtex4}
\usepackage{graphicx}

\begin{document}

\title{Directed update for the Stochastic Green Function algorithm}

\author{V.G.~Rousseau}
\affiliation{Instituut-Lorentz, LION, Universiteit Leiden, Postbus 9504, 2300 RA Leiden, The Netherlands}

\begin{abstract}
In a recent publication we have presented the stochastic Green function (SGF)
algorithm, which has the properties of being general and easy
to apply to any lattice Hamiltonian of the form $\hat\mathcal H=\hat\mathcal V-\hat\mathcal T$,
where $\hat\mathcal V$ is diagonal in the chosen occupation number basis and $\hat\mathcal T$ 
has only positive matrix elements. We propose here a modified version of the
update scheme that keeps the simplicity and generality of the original SGF algorithm, and
enhances significantly its efficiency.
\end{abstract}

\maketitle

\section{Introduction}

Monte Carlo methods \cite{MonteCarlo} appeared about sixty years ago with the need
to evaluate numerical values for various complex problems. These
methods evolved and were applied early to quantum problems, thus
putting within reach exact numerical solutions to non-trivial quantum
problems \cite{QMC1,QMC2,QMC3,Batrouni1992}. Many improvements of these methods followed, avoiding
critical slowing down near phase transitions and allowing to work
directly in the continuous imaginary time limit \cite{QMC4,QMC5,QMC6,QMC7,QMC8,Sandvik,Prokofev,Rigol03}. In recent
years, interest in methods that work in the canonical ensemble with
global updates yet allow access to Green functions has intensified \cite{VanHoucke06,RousseauSGF}. However, a method that works well for a given Hamiltonian
often needs major modifications for another. For example, the addition
of a 4-site ring exchange term in the bosonic Hubbard model required
special developments for a treatment by the stochastic series
expansion algorithm \cite{Sandvik2002}, as well as by the wordline algorithm \cite{Rousseau2005}. This can result in long delays. It is, therefore, advantageous
to have at one's disposal an algorithm that can be applied to a very
wide class of Hamiltonians without requiring any changes. In a recent
publication \cite{RousseauSGF}, the stochastic Green function (SGF) algorithm was
presented, which meets this goal. The algorithm can be applied to any lattice Hamiltonian of the form
\begin{equation}
  \label{Ham} \hat\mathcal H=\hat\mathcal V-\hat\mathcal T,
\end{equation}
where $\hat\mathcal V$ is diagonal in the chosen occupation number basis and $\hat\mathcal T$ has only positive matrix elements.
This includes all kinds of systems that can be treated by other methods presented in ref.\cite{Batrouni1992,Sandvik,Prokofev,Rigol03,VanHoucke06},
for instance Bose-Hubbard models with or without a trap, Bose-Fermi mixtures in one dimension, Heisenberg models...
In particular Hamiltonians for which the non-diagonal part $\hat\mathcal T$ is non-trivial (the eigen-basis is unknown)
are easily treated, such as the Bose-Hubbard model with ring exchange \cite{Sandvik2002,Rousseau2005}, or multi-species
Hamiltonians in which a given species can be turned into another one (see eq.(\ref{TwoSpecies}) and fig. \ref{Density} and \ref{Momentum} for a concrete example).
Systems for which it is not possible to find a basis in which $\hat\mathcal V$ is diagonal and $\hat\mathcal T$ has only positive matrix elements
are said to have a "sign problem", which usually arises with fermionic and frustrated systems. As other QMC methods, the
SGF algorithm does not solve this problem.

The algorithm allows to measure several quantities of interest, such as the energy, the local density, local compressibility,
density-density correlation functions... In particular the winding is sampled and gives access to the superfluid density.
Equal-time n-body Green functions are probably the most interesting quantities that can be measured by the algorithm,
by giving access to momentum distribution functions which allow direct comparisons with experiments. All details on
measurements are given in ref.\cite{RousseauSGF}.

In addition the algorithm has the property of being easy to code, due in part to a simple update scheme in which
all moves are accepted with a probability of 1. Despite of such generality and simplicity, the algorithm might suffer from a reduced efficiency, compared
to other algorithms in situations where they can be applied.

The purpose of this paper is to present a "directed" update scheme that (i) keeps the simplicity and generality of the original SGF algorithm, and (ii) enhances
its efficiency by improving the sampling over the imaginary time axis.
While the SGF algorithm is not intended to compete with the speed of other algorithms, the improvment resulting
from the directed update scheme is remarkable (see section V). But what makes the strength of the SGF method
is that it allows to simulate Hamiltonians that cannot be treated by other methods or that would require
special developments (see eq.(\ref{TwoSpecies}) for a concrete example).
The paper is organized as follows: We introduce in section II the notations and definitions
used in ref.\cite{RousseauSGF}. In section III, we propose a simplification of the update scheme used in the original SGF algorithm, and determine how to satisfy detailed balance.
A generalization of the simplified update scheme is presented in section IV, which constitutes the directed updated scheme. Finally section V shows how to determine
the introduced optimization parameters, and presents some tests of the algorithm and a comparison with the original version.

\section{Definitions and notations}
In this section, we recall the expression of the "Green operator" introduced in
the SGF algorithm, and the extended partition function which is considered. Although
not required for understanding this paper,
we refer the reader to ref.\cite{RousseauSGF} for full details on the
algorithm. As many QMC algorithms, the SGF algorithm samples the partition
function
\begin{equation}
  \label{PartitionFunction} \mathcal Z(\beta)=\textrm{Tr }e^{-\beta\hat\mathcal H}.
\end{equation}
The algorithm has the property of working in the canonical ensemble. In order to define the Green operator, we first define the "normalized"
creation and annihilation operators,
\begin{equation}
  \label{NormalizedOperators} \hat\mathcal A^\dagger=a^\dagger\frac{1}{\sqrt{\hat n+1}} \hspace{1cm} \hat\mathcal A=\frac{1}{\sqrt{\hat n+1}}a,
\end{equation}
where $a^\dagger$ and $a$ are the usual creation and annihilation operators
of bosons, and $\hat n=a^\dagger a$ is the number operator. From
(\ref{NormalizedOperators}) one can show the following relations for any state $\big|n\big\rangle$
in the occupation number representation,
\begin{equation}
  \hat\mathcal A^\dagger\big|n\big\rangle=\big|n+1\big\rangle \hspace{1cm} \hat\mathcal A\big|n\big\rangle=\big|n-1\big\rangle,
\end{equation}
with the particular case $\hat\mathcal A\big|0\big\rangle=0$. Appart from
this exception, the operators $\hat\mathcal A^\dagger$ and $\hat\mathcal A$
change a state $\big|n\big\rangle$ by respectively creating and annihilating
one particle, but
they do not change the norm of the state.

Using the notation $\big\lbrace i_p|j_q\big\rbrace$ to denote two subsets
of site indices $i_1,i_2,\cdots,i_p$ and $j_1,j_2,\cdots,j_q$ with the
constraint that all indices in subset $i$ are different from the indices in subset $j$
(but several indices in one subset may be equal), we define the Green operator
$\hat\mathcal G$ by
\begin{equation}
  \label{GreenOperator} \hat\mathcal G=\sum_{p=0}^{+\infty}\sum_{q=0}^{+\infty}g_{pq}\sum_{\big\lbrace i_p|j_q\big\rbrace}\prod_{k=1}^p\hat\mathcal A_{i_k}^\dagger \prod_{l=1}^q\hat\mathcal A_{j_l},
\end{equation}
where $g_{pq}$ is a matrix that depends on the application of the algorithm \cite{RousseauSGF}.
In order to sample the partition function (\ref{PartitionFunction}), an
extended partition function $\mathcal Z(\beta,\tau)$ is considered by
breaking up the propagator $e^{-\beta\hat\mathcal H}$, and introducing
the Green operator between the broken parts,
\begin{equation}
  \label{ExtendedSpace} Z(\beta,\tau)=\textrm{Tr }e^{-(\beta-\tau)\hat\mathcal H}\hat\mathcal G e^{-\tau\hat\mathcal H}.
\end{equation}
Defining the time dependant operators $\hat\mathcal T(\tau)$ and $\hat\mathcal G(\tau)$,
\begin{equation}
  \label{TimeRepresentation} \hat\mathcal T(\tau)=e^{\tau\hat\mathcal V}\hat\mathcal T e^{-\tau\hat\mathcal V} \hspace{0.5cm} \hat\mathcal G(\tau)=e^{\tau\hat\mathcal V}\hat\mathcal G e^{-\tau\hat\mathcal V},
\end{equation}
and working in the occupation number basis in which $\hat\mathcal V$ is
diagonal, the extended partition function takes the form
\begin{eqnarray}
  \nonumber && \!\!\!\!\!\!\!\mathcal Z(\beta,\tau)\!\!=\!\!\! \sum_{n\geq 0}\int_{0<\tau_1<\cdots<\tau_n<\beta} \hspace{-2cm}\big\langle\psi_0\big|e^{-\beta\mathcal V}\hat\mathcal T(\tau_n)\big|\psi_{n-1}\big\rangle\big\langle\psi_{n-1}\big|\hat\mathcal T(\tau_{n-1})\big|\psi_{n-2}\big\rangle\\
  \label{ExtendedPartitionFunction} && \!\!\!\!\!\!\!\times \cdots \big\langle\psi_{L+1}\big|\hat\mathcal T(\tau_L)\big|\psi_L\big\rangle \big\langle\psi_L\big|\hat\mathcal G(\tau)\big|\psi_R\big\rangle \big\langle\psi_R\big|\hat\mathcal T(\tau_R)\big|\psi_{R-1}\big\rangle \\
  \nonumber && \!\!\!\!\!\!\!\times \cdots \big\langle\psi_2\big|\hat\mathcal T(\tau_2)\big|\psi_{1}\big\rangle \big\langle\psi_1\big|\hat\mathcal T(\tau_1)\big|\psi_0\big\rangle d\tau_1\cdots d\tau_n,
\end{eqnarray}
where the sum $\sum_{n\geq 0}$ implicitly runs over complete sets of states $\big\lbrace\big|\psi_k\big\rangle\big\rbrace$.
We will systematically use the labels $L$ and $R$
to denote the states appearing on the left and the right of the Green
operator, and use the notation $V_k$ to denote the diagonal energy $\big\langle\psi_k\big|\hat\mathcal V\big|\psi_k\big\rangle$.
We will also denote by $\tau_L$ and $\tau_R$ the time indices
of the $\hat\mathcal T$ operators appearing on the left and the right
of $\hat\mathcal G$.

As a result, the extended partition function is a sum over all possible
configurations, each being determined by a set of time indices $\tau_1,\tau_2,\cdots,\tau_R,\tau,\tau_L,\cdots,\tau_n$
and a set of states $\big|\psi_0\big\rangle$, $\big|\psi_1\big\rangle$,
$\cdots\big|\psi_R\big\rangle$,$\big|\psi_L\big\rangle$,
$\cdots\big|\psi_{n-1}\big\rangle$. The algorithm consists in updating
those configurations by making use of the Green operator. Assuming that the Green
operator is acting at time $\tau$, it can "create" a $\hat\mathcal T$ operator (that is to say a $\hat\mathcal T$
operator can be inserted in the operator string)
at the same time, thus introducing a new intermediate state, then it can be shifted to a different time.
While shifting, any $\hat\mathcal T$ operator encountered by the Green
operator is "destroyed" (that is to say removed from the operator string). Assuming a left (or right) move, creating an operator will update the state $\big|\psi_R\big\rangle$
(or  $\big|\psi_L\big\rangle$), while destroying will update the state
$\big|\psi_L\big\rangle$ (or  $\big|\psi_R\big\rangle$). When a diagonal configuration of the Green operator occurs, $\big|\psi_L\big\rangle=\big|\psi_R\big\rangle$,
such a configuration associated to the extended partition function (\ref{ExtendedPartitionFunction}) is also a configuration associated to the
partition function (\ref{PartitionFunction}). Measurements can be done when this occurs (see ref.\cite{RousseauSGF} for details on measurements).

Next section presents a simple update scheme that meets the requirements
of ergodicity and detailed balance.

\section{Simplified update scheme}
Before introducing the directed update, we start by simplifying the
update scheme used in the original SGF algorithm.

\subsection{The update scheme}
\begin{figure}[h]
  \centerline{\includegraphics[width=0.45\textwidth]{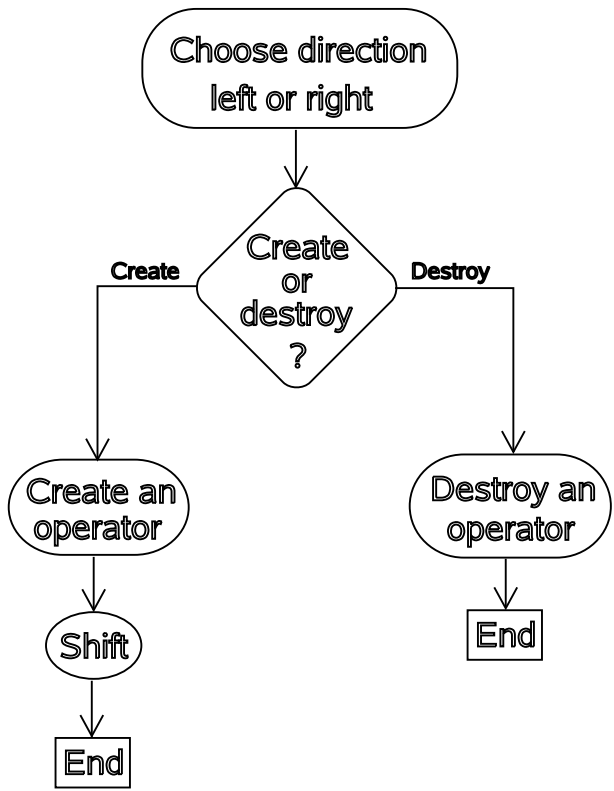}}
  \caption
    {
      The simplified update scheme. See text for details.
    }
  \label{SimplfiedUpdateScheme}
\end{figure}

We will assume in the following that a left move of the
Green operator is chosen. In the original version, the Green operator
$\hat\mathcal G(\tau)$ can choose to create or not on its right a $\hat\mathcal T$ operator at time $\tau$.
Then a time shift $\Delta\tau$ to the left is chosen for the Green operator with an exponential
distribution in the range $[0;+\infty[$. If an operator is encountered while shifting
the Green operator, then the operator is destroyed and the move stops there.
As a result, four possible situations can occur during one move:
\begin{enumerate}
  \item No creation, shift, no destruction.
  \item Creation, shift, no destruction.
  \item No creation, shift, destruction.
  \item Creation, shift, destruction.
\end{enumerate}
It appears that the first possibility "no creation, no destruction" is
actually useless, since no change is performed in the operator string.
The idea is to get rid of this possibility by forcing the Green operator
to destroy an operator if no creation is chosen. A further simplification
can be done by noticing that the last possibility "creation, destruction"
is not necessary for the ergodicity of the algorithm, and can be avoided by
restricting the range of the time shift after having created an operator. Therefore we replace
the original update scheme by the following: We assume that the Green
operator is acting at time $\tau$ and that the operator on its left is acting at
time $\tau_L$. The Green operator $\hat\mathcal G(\tau)$
chooses to create or not an operator on its right at time $\tau$. If
creation is chosen, then a time shift $\Delta\tau$ of the Green operator is chosen to the left in the range $[0;\tau_L-\tau[$, with the probability distribution defined below.
If no creation is chosen, then the Green operator is directly
shifted to the operator on its left at time $\tau_L$, and the operator is
destroyed. As a result only two possibilities have to be considered:
\begin{enumerate}
  \item Creation, shift.
  \item Shift, destruction.
\end{enumerate} 
Figure \ref{SimplfiedUpdateScheme} shows the associated organigram.
Section III.B explains how detailed balance can be satisfied with
this simplified update scheme.

\subsection{Detailed balance}
When updating the configurations according to the chosen update scheme, we need to generate
different transitions from initial to final states with probabilities that satisfy detailed balance.
In this section we propose a choice for these probabilities, and determine the corresponding
acceptance factors. We denote the probability of the initial (final) configuration by $P_i$ ($P_f$). We denote
by $S_{i\to f}$ the probability of the transition from configuration $i$ to configuration $f$, and by
$S_{f\to i}$ the probability of the reverse transition. Finally we denote by $A_{i\to f}$ the acceptance rate
of the transition from $i$ to $f$, and by $A_{f\to i}$ the acceptance rate of the reverse transition. The
detailed balance can be written as
\begin{equation}
  \label{DetailedBalance} P_i S_{i\to f} A_{i\to f}=P_f S_{f\to i} A_{f\to i}.
\end{equation}
We will make use of the Metropolis solution \cite{Metropolis},
\begin{equation}
  \label{Metropolis1} A_{i\to f}=\min(1,q)
\end{equation}
with
\begin{equation}
  \label{Metropolis2} q=\frac{P_f S_{f\to i}}{P_i S_{i\to f}}.
\end{equation}
We will use primed (non-primed) labels for states and time indices to denote final (initial) configurations.

\subsubsection{Creation, shift}
We consider here the case where a left move is chosen, an operator is created on the right of
the Green operator at time $\tau$, and a new state is chosen. Then a time shift to the left
is chosen for the Green operator in the range $[0,\tau_L^\prime-\tau_R^\prime[$.
It is important to note that $\tau_L^\prime$ and $\tau_R^\prime$ correspond to the time indices
of the operators appearing on the left and the right of the Green operator after the new operator
has been inserted, that is to say at the moment where the time shift needs to be performed. Thus we have
$\tau_L^\prime=\tau_L$ and $\tau_R^\prime=\tau$.

The probability of the initial configuration is the Boltzmann weight appearing
in the extended partition function (\ref{ExtendedPartitionFunction}):
\begin{eqnarray}
  \nonumber P_i   & \propto & \big\langle\psi_L\big|\hat\mathcal G(\tau)\big|\psi_R\big\rangle \\
  \label{Initial} & \propto & e^{\tau V_L}\big\langle\psi_L\big|\hat\mathcal G\big|\psi_R\big\rangle e^{-\tau V_R}
\end{eqnarray}
The probability of the final configuration takes the form:
\begin{eqnarray}
  \nonumber P_f & \propto & \big\langle\psi_L^\prime\big|\hat\mathcal G(\tau^\prime)\big|\psi_R^\prime\big\rangle\big\langle\psi_R^\prime\big|\hat\mathcal T(\tau_R^\prime)\big|\psi_{R-1}^\prime\big\rangle \\
  \label{Final} & \propto & e^{\tau^\prime V_L^\prime}\big\langle\psi_L^\prime\big|\hat\mathcal G\big|\psi_R^\prime\big\rangle e^{-(\tau^\prime-\tau_R^\prime)V_R^\prime} \big\langle\psi_R^\prime\big|\hat\mathcal T\big|\psi_{R-1}^\prime\big\rangle e^{-\tau_R^\prime V_{R-1}^\prime}
\end{eqnarray}
It is important here to realize that the Green operator only inserted on its right the operator
$\big|\psi_R^\prime\big\rangle\big\langle\psi_R^\prime\big|\hat\mathcal T(\tau_R^\prime)$, before being shifted
from $\tau_R^\prime$ to $\tau^\prime$. Therefore we have the equalities $\big\langle\psi_L^\prime\big|=\big\langle\psi_L\big|$,
$\big|\psi_{R-1}^\prime\big\rangle=\big|\psi_R\big\rangle$, $V_L^\prime=V_L$, and $V_{R-1}^\prime=V_R$.

The probability $S_{i\to f}$ of the transition from the initial configuration to
the final configuration is the probability $P(\leftarrow)$ of a left move,
times the probability $P_\leftarrow^\dagger(\tau)$ of a creation, times the probability
$P_\leftarrow(\psi_R^\prime)$ to choose the new state $\psi_R^\prime$, times the probability
$P_\leftarrow^{L^\prime R^\prime}(\tau^\prime-\tau_R^\prime)$ to shift the Green operator
by $\tau^\prime-\tau_R^\prime$, knowing that the states on the left and the right of the Green operator at the moment of the shift
are $\psi_{L}^\prime$ and $\psi_{R}^\prime$:
\begin{equation}
  S_{i\to f}=P(\leftarrow)P_\leftarrow^\dagger(\tau)P_\leftarrow(\psi_R^\prime)P_\leftarrow^{L^\prime R^\prime}(\tau^\prime-\tau_R^\prime)
\end{equation}
The probability of the reverse transition
is simply the probability $P(\rightarrow^\prime)$ of a right move,
times the probability of no creation, $1-P_\rightarrow^\dagger(\tau^\prime)$:
\begin{equation}
  S_{f\to i}=P(\rightarrow^\prime)\big[1-P_\rightarrow^\dagger(\tau^\prime)\big]
\end{equation}
From the original version of the SGF algorithm, we know that choosing the
time shift with an exponential distribution is a good choice, because it
cancels the exponentials appearing in the probabilities of the initial (\ref{Initial})
and final (\ref{Final}) configurations, avoiding exponentially small acceptance
factors. However a different normalization must be used here, since the
time shift is chosen in the range $[0;\tau_L^\prime-\tau_R^\prime[$ instead of $[0;+\infty[$.
The suitable solution is:
\begin{equation}
  \label{ExponentialDistribution} P_\leftarrow^{L^\prime R^\prime}(\Delta\tau)=\frac{(V_R^\prime-V_L^\prime)e^{-\Delta\tau(V_R^\prime-V_L^\prime)}}{1-e^{-(\tau_L^\prime-\tau_R^\prime)(V_R^\prime-V_L^\prime)}}
\end{equation}
It is straightforward to check that the above probability is correctly
normalized and well-defined for any real value of $V_R^\prime-V_L^\prime$, the particular
case $V_L^\prime=V_R^\prime$ reducing to the uniform distribution $P(\Delta\tau)=1/(\tau_L^\prime-\tau_R^\prime)$
(note that $\tau_L^\prime-\tau_R^\prime$ is always a positive number).
For the probability $P_\leftarrow(\psi_R^\prime)$ to choose the new state $\psi_R^\prime$, the convenient
solution is the same as in the original version:
\begin{equation}
  P_\leftarrow(\psi_R^\prime)=\frac{\big\langle\psi_L\big|\hat\mathcal G\big|\psi_R^\prime\big\rangle\big\langle\psi_R^\prime|\hat\mathcal T\big|\psi_R\big\rangle}{\big\langle\psi_L\big|\hat\mathcal G\hat\mathcal T\big|\psi_R\big\rangle}
\end{equation}
Putting everything together, the acceptance factor (\ref{Metropolis2}) becomes
\begin{eqnarray}
  \nonumber q_\leftarrow^c &=&      \frac{\big\langle\psi_L\big|\hat\mathcal G\hat\mathcal T\big|\psi_R\big\rangle}{\big\langle\psi_L\big|\hat\mathcal G\big|\psi_R\big\rangle P(\leftarrow)P_\leftarrow^\dagger(\tau)} \\
                           &\times& \frac{P(\rightarrow^\prime)\big[1-P_\rightarrow^\dagger(\tau^\prime)\big]\big[1-e^{-(\tau_L^\prime-\tau_R^\prime)(V_R^\prime-V_L^\prime)}\big]}{V_R^\prime-V_L^\prime},
\end{eqnarray}
where we have used the notation $q_\leftarrow^c$ to emphasize that this acceptance factor
corresponds to a creation. It is also important for the remaining of this
paper to note that $q_\leftarrow^c$ is written as a quantity that depends on the
initial configuration, times a quantity that depends on the final configuration.

\subsubsection{Shift, destruction}
We consider here the case where a left move is chosen, and the operator
on the left of the Green operator is destroyed. This move corresponds to
the inverse of the above "creation, shift" move. Thus, the corresponding
acceptance factor $q_\leftarrow^d$ is obtained by inverting the acceptance factor
$q_\leftarrow^c$, exchanging the initial time $\tau$ and final time $\tau^\prime$, and
switching the direction. However $\tau_L-\tau_R$ represents an absolute time shift, so $\tau_L$ and $\tau_R$ do not have to be exchanged. We get
\begin{eqnarray}
  \nonumber q_\leftarrow^d &=&      \frac{V_L-V_R}{P(\leftarrow)\big[1-P_\leftarrow^\dagger(\tau)\big]\big[1-e^{-(\tau_L-\tau_R)(V_L-V_R)}\big]} \\
                           &\times& \frac{\big\langle\psi_L^\prime\big|\hat\mathcal G\big|\psi_R^\prime\big\rangle P(\rightarrow^\prime)P_\rightarrow^\dagger(\tau^\prime)}{\big\langle\psi_L^\prime\big|\hat\mathcal T\hat\mathcal G\big|\psi_R^\prime\big\rangle},
\end{eqnarray}
which is written as a quantity that depends on the initial configuration,
times a quantity that depends on the final configuration.

\subsubsection{Simplification of the acceptance factors}
We will use here the short notation $\big\langle\hat\mathcal G\big\rangle$, $\big\langle\hat\mathcal G\hat\mathcal T\big\rangle$, and $\big\langle\hat\mathcal T\hat\mathcal G\big\rangle$
to denote respectively the quantities $\big\langle\psi_L\big|\hat\mathcal G\big|\psi_R\big\rangle$, $\big\langle\psi_L\big|\hat\mathcal G\hat\mathcal T\big|\psi_R\big\rangle$,
and $\big\langle\psi_L\big|\hat\mathcal T\hat\mathcal G\big|\psi_R\big\rangle$.
As in ref. \cite{RousseauSGF}, we have some freedom for the choice
of the probabilities of choosing a left or right move, $P(\leftarrow)$ and
$P(\rightarrow)$, and the probabilities of creation $P_\leftarrow^\dagger(\tau)$ and
$P_\rightarrow^\dagger(\tau)$. A suitable choice for those probabilities
can be done in order to accept all moves, resulting in an appreciable simplification
of the algorithm. For this purpose, we impose the acceptance factor
$q_\leftarrow^c$ (or  $q_\rightarrow^c$) to be equal to the acceptance factor $q_\leftarrow^d$ (or  $q_\rightarrow^d$).
This allows to determine the probabilities $P_\leftarrow^\dagger(\tau)$ and
$P_\rightarrow^\dagger(\tau)$,
\begin{eqnarray}
  && P_\leftarrow^\dagger(\tau)=\frac{\big\langle\hat\mathcal G\hat\mathcal T\big\rangle}{\big\langle\hat\mathcal G\hat\mathcal T\big\rangle+\big\langle\hat\mathcal G\big\rangle\frac{(V_L-V_R)}{1-e^{-(\tau_L-\tau_R)(V_L-V_R)}}} \\
  && P_\rightarrow^\dagger(\tau)=\frac{\big\langle\hat\mathcal T\hat\mathcal G\big\rangle}{\big\langle\hat\mathcal T\hat\mathcal G\big\rangle+\big\langle\hat\mathcal G\big\rangle\frac{(V_R-V_L)}{1-e^{-(\tau_L-\tau_R)(V_R-V_L)}}},
\end{eqnarray}
and the acceptance factors $q_\leftarrow^c=q_\leftarrow^d$ and $q_\rightarrow^c=q_\rightarrow^d$
take the form
\begin{equation}
  q_\leftarrow=\frac{P(\rightarrow^\prime)r_\leftarrow(\tau)}{P(\leftarrow)r_\rightarrow(\tau^\prime)} \hspace{1cm} q_\rightarrow=\frac{P(\leftarrow^\prime)r_\rightarrow(\tau)}{P(\rightarrow)r_\leftarrow(\tau^\prime)},
\end{equation}
with
\begin{eqnarray}
  && r_\leftarrow(\tau)=\frac{\big\langle\hat\mathcal G\hat\mathcal T\big\rangle}{\big\langle\hat\mathcal G\big\rangle}+\frac{V_L-V_R}{1-e^{-(\tau_L-\tau_R)(V_L-V_R)}} \\
  && r_\rightarrow(\tau)=\frac{\big\langle\hat\mathcal T\hat\mathcal G\big\rangle}{\big\langle\hat\mathcal G\big\rangle}+\frac{V_R-V_L}{1-e^{-(\tau_L-\tau_R)(V_R-V_L)}}.
\end{eqnarray}
Finally we can impose the acceptance factors $q_\leftarrow$ and $q_\rightarrow$
to be equal. This implies
\begin{equation}
  P(\leftarrow)=\frac{r_\leftarrow(\tau)}{r_\leftarrow(\tau)+r_\rightarrow(\tau)} \hspace{0.5cm} P(\rightarrow)=\frac{r_\rightarrow(\tau)}{r_\leftarrow(\tau)+r_\rightarrow(\tau)}.
\end{equation}
Defining $R(\tau)=r_\leftarrow(\tau)+r_\rightarrow(\tau)$, we are left with a single acceptance factor,
\begin{equation}
  q=\frac{R(\tau)}{R(\tau^\prime)},
\end{equation}
which is independent of the chosen direction, and independent of the nature of the move (creation or destruction).
Thus all moves can be accepted by making use of a proper reweighting, as explained in ref. \cite{RousseauSGF}.
The appendix shows how to generate random numbers with the appropriate exponential distribution (\ref{ExponentialDistribution}).
\subsection{Discussion}
Although the above simplified update scheme works, it turns out to have a poor efficiency.
This is because of a
lack of "directionality": The Green operator has, in average, a probability
of $1/2$ to choose a left move or a right move. Therefore the Green operator
propagates along the operator string like a "drunk man", with a diffusion-like
law. The basic creation and destruction processes correspond to the steps of the random walk.

This suggests that the efficiency of the update scheme can be improved
if one can force the Green operator to move in the same direction for
several iterations. Next section presents a modified version of the simplified
update scheme, which allows to control the mean length of the steps of the random walk, that is to say the mean number of creations and destructions in a given
direction. The proposed directed update scheme can be considered 
analogous to the "directed loop update" used in the stochastic series expansion
algorithm \cite{Sandvik,Syljuasen}, which prevents a worm from going backwards.
However the connection should not be pushed too far. Indeed the picture of a worm whose
head is evolving both in space and imaginary time accross vertices is obvious in a loop algorithm. In
such algorithm, a creation (or an annihilation) operator which is represented by the head of a worm is propagated both
in space and imaginary time, while an annihilation (or a creation) operator represented by the tail of the worm remains at rest.
The loop ends when the head of the worm bites the tail.

Such a worm picture is not obvious in the SGF algorithm: Instead of single creation or annihilation operators, it is the full Green operator over the whole space
that is propagated only in imaginary time. This creates open worldlines, thus introducing discontinuities. These
discontinuities increase or decrease while propagating in imaginary time.
All open ends of the worldlines are localized at the same imaginary time index.
Therefore it is actually not possible to draw step by step a worm whose head is evolving in space and imaginary time until
it bites its tail.

\section{Directed update scheme}
We present in this section a directed update scheme which is obtained
by modifying slightly the simplified update scheme, thus keeping the simplicity
and generality of the algorithm.

\subsection{The update scheme}
\begin{figure}[h]
  \centerline{\includegraphics[width=0.45\textwidth]{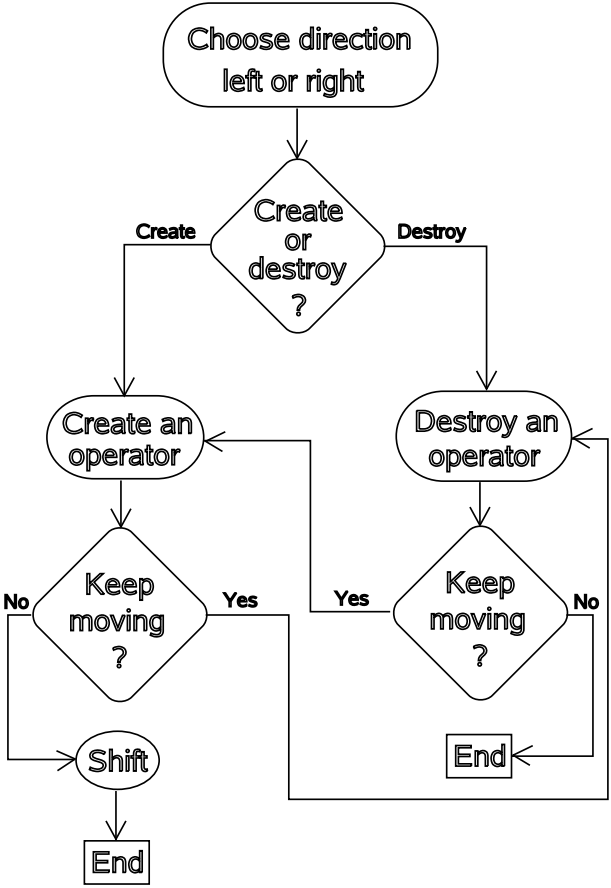}}
  \caption
    {
      The directed update scheme. See text for details.
    }
  \label{DirectedUpdateScheme}
\end{figure}

Assuming that a left move is chosen, the Green operator chooses between
starting the move by a creation or a destruction. After having created (or  destroyed)
an operator, the Green operator can choose to keep moving in the same direction and destroy (or  create) with a probability
$P_\leftarrow^{kd}$ (or  $P_\leftarrow^{kc}$),
or to stop. If it keeps moving, then a destruction (or  creation) occurs, and
the Green operator can choose to keep moving and create (or  destroy) with a probability
$P_\leftarrow^{kc}$ (or  $P_\leftarrow^{kd}$)... and so on, until it
decides to stop. If the last action of the move is a creation, then a
time shift is chosen. The organigram is represented in Figure \ref{DirectedUpdateScheme}.

\subsection{Detailed balance}
In order to satisfy detailed balance, in addition to the acceptance
factors $q_\leftarrow^c$ and $q_\leftarrow^d$, we need to determine
new acceptance factors of the form $q_\leftarrow^{cdcdcdc\cdots}$ and
$q_\leftarrow^{dcdcdcd\cdots}$. We first determine the new expressions of $q_\leftarrow^c$ and $q_\leftarrow^d$
resulting from the directed update scheme. For $q_\leftarrow^c$, the previous probability
$S_{i\to f}$ has to be multiplied by the probability to stop the move after having created,
$1-P_\leftarrow^{kd}(\tau^\prime)$. The previous probability $S_{f\to i}$
has to be multiplied by the probability to stop the move after having destroyed,
$1-P_\rightarrow^{kc}(\tau)$. We get for $q_\leftarrow^c$ and $q_\leftarrow^d$
the new expressions:
\begin{eqnarray}
  \nonumber q_\leftarrow^c &=&      \frac{\big\langle\psi_L\big|\hat\mathcal G\hat\mathcal T\big|\psi_R\big\rangle\big[1-P_\rightarrow^{kc}(\tau)\big]}{\big\langle\psi_L\big|\hat\mathcal G\big|\psi_R\big\rangle P(\leftarrow)P_\leftarrow^\dagger(\tau)} \\
                           &\times& \frac{P(\rightarrow^\prime)\big[1-P_\rightarrow^\dagger(\tau^\prime)\big]\big[1-e^{-(\tau_L^\prime-\tau_R^\prime)(V_R^\prime-V_L^\prime)}\big]}{\big[1-P_\leftarrow^{kd}(\tau^\prime)\big]\big(V_R^\prime-V_L^\prime\big)} \\
  \nonumber q_\leftarrow^d &=&      \frac{\big[1-P_\rightarrow^{kd}(\tau)\big]\big(V_L-V_R\big)}{P(\leftarrow)\big[1-P_\leftarrow^\dagger(\tau)\big]\big[1-e^{-(\tau_L-\tau_R)(V_L-V_R)}\big]} \\
                           &\times& \frac{\big\langle\psi_L^\prime\big|\hat\mathcal G\big|\psi_R^\prime\big\rangle P(\rightarrow^\prime)P_\rightarrow^\dagger(\tau^\prime)}{\big\langle\psi_L^\prime\big|\hat\mathcal T\hat\mathcal G\big|\psi_R^\prime\big\rangle\big[1-P_\leftarrow^{kc}(\tau^\prime)\big]},
\end{eqnarray}
\subsubsection{Creation, destruction}
We consider here the case where a left move is chosen, an operator is created on the right of
the Green operator, and a new state is chosen. Then the operator on the
left of the Green operator is destroyed. Using the superscripts $a,b,c,\cdots$ to denote intermediate configurations
between initial and final configurations, the sequence is the following
\begin{enumerate}
  \item $P_i\propto\big\langle\psi_{L+1}\big|\hat\mathcal T(\tau_L)\big|\psi_L\big\rangle\big\langle\psi_L\big|\hat\mathcal G(\tau)\big|\psi_R\big\rangle$
  \item $\big\langle\psi_{L+1}^a\big|\hat\mathcal T(\tau_L^a)\big|\psi_L^a\big\rangle\big\langle\psi_L^a\big|\hat\mathcal G(\tau^a)\big|\psi_R^a\big\rangle\big\langle\psi_R^a\big|\hat\mathcal T(\tau_R^a)\big|\psi_{R-1}^a\big\rangle$
  \item $P_f\propto\big\langle\psi_L^\prime\big|\hat\mathcal G(\tau^\prime)\big|\psi_R^\prime\big\rangle\big\langle\psi_R^\prime\big|\hat\mathcal T(\tau_R^\prime)\big|\psi_{R-1}^\prime\big\rangle$,
\end{enumerate}
where we have $\big\langle\psi_{L+1}\big|=\big\langle\psi_{L+1}^a\big|=\big\langle\psi_L^\prime\big|$, $\big|\psi_R\big\rangle=\big|\psi_{R-1}^a\big\rangle=\big|\psi_{R-1}^\prime\big\rangle$, $\big|\psi_L\big\rangle\big\langle\psi_L\big|=\big|\psi_L^a\big\rangle\big\langle\psi_L^a\big|$, and
$\big|\psi_R^a\big\rangle\big\langle\psi_R^a\big|=\big|\psi_R^\prime\big\rangle\big\langle\psi_R^\prime\big|$. The probability of the transition from the initial
configuration to the final configuration is the probability $P(\leftarrow)$ to choose
a left move, times the probability $P_\leftarrow^\dagger(\tau)$ to create
an operator at time $\tau$, times the probability $P_\leftarrow(\psi_R^a)$ to choose
the new state $\psi_R^a$, times the probability $P_\leftarrow^{kd}(a)$ to
keep moving and destroy, times the probability $1-P_\leftarrow^{kc}(\tau^\prime)$ to stop the move after
having destroyed:
\begin{equation}
  S_{i\to f}=P(\leftarrow)P_\leftarrow^\dagger(\tau)P_\leftarrow(\psi_R^a)P_\leftarrow^{kd}(a)\big[1-P_\leftarrow^{kc}(\tau^\prime)\big]
\end{equation}
The probability of the reverse move is exactly symmetric:
\begin{equation}
  S_{f\to i}=P(\rightarrow^\prime)P_\rightarrow^\dagger(\tau^\prime)P_\rightarrow(\psi_L^a)P_\rightarrow^{kd}(a)\big[1-P_\rightarrow^{kc}(\tau)\big]
\end{equation}
It is important to notice that, when in the intermediate configuration $a$, the 
time $\tau_L^a$ of the operator to the left of the Green operator is equal
to $\tau_L$, and the time $\tau_R^a$ of the operator to the right of the
Green operator is equal to $\tau$. Thus the acceptance factor takes the form
\begin{eqnarray}
  \nonumber q_\leftarrow^{cd} & =    & \frac{\big\langle\psi_L\big|\hat\mathcal G\hat\mathcal T\big|\psi_R\big\rangle\big[1-P_\rightarrow^{kc}(\tau)\big]}{\big\langle\psi_L\big|\hat\mathcal G\big|\psi_R\big\rangle P(\leftarrow)P_\leftarrow^\dagger(\tau)} \\
  \nonumber                   &\times& \frac{e^{-\big(\tau_L^a-\tau_R^a\big)V_R^a}P_\rightarrow^{kd}(a)}{e^{-\big(\tau_L^a-\tau_R^a\big)V_L^a}P_\leftarrow^{kd}(a)} \\
                              &\times& \frac{\big\langle\psi_L^\prime\big|\hat\mathcal G\big|\psi_R^\prime\big\rangle P(\rightarrow^\prime)P_\rightarrow^\dagger(\tau^\prime)}{\big\langle\psi_L^\prime\big|\hat\mathcal T\hat\mathcal G\big|\psi_R^\prime\big\rangle\big[1-P_\leftarrow^{kc}(\tau^\prime)\big]},
\end{eqnarray}
and is written as a quantity that depends on the initial configuration, times
a quantity that depends on the intermediate configuration $a$, times a
quantity that depends on the final configuration. It is useful for the
remaining of the paper to define the intermediate acceptance factor,
\begin{equation}
  q_\leftarrow^{c-d}(a)=\frac{e^{-\big(\tau_L^a-\tau_R^a\big)V_R^a}P_\rightarrow^{kd}(a)}{e^{-\big(\tau_L^a-\tau_R^a\big)V_L^a}P_\leftarrow^{kd}(a)}.
\end{equation}

\subsubsection{Destruction, creation}
We consider here the case where a left move is chosen, the operator on the left
of the Green operator is destroyed, then an operator is created on its
right, and a new state is chosen. Finally a time shift is chosen. The sequence of configurations
is the following
\begin{enumerate}
  \item $P_i\propto\big\langle\psi_{L+1}\big|\hat\mathcal T(\tau_L)\big|\psi_L\big\rangle\big\langle\psi_L\big|\hat\mathcal G(\tau)\big|\psi_R\big\rangle$
  \item $\big\langle\psi_L^a\big|\hat\mathcal G(\tau^a)\big|\psi_R^a\big\rangle$
  \item $P_f\propto\big\langle\psi_L^\prime\big|\hat\mathcal G(\tau^\prime)\big|\psi_R^\prime\big\rangle\big\langle\psi_R^\prime\big|\hat\mathcal T(\tau_R^\prime)\big|\psi_{R-1}^\prime\big\rangle$,
\end{enumerate}
where we have $\big\langle\psi_{L+1}\big|=\big\langle\psi_L^a\big|=\big\langle\psi_L^\prime\big|$, and $\big|\psi_R\big\rangle=\big|\psi_R^a\big\rangle=\big|\psi_{R-1}^\prime\big\rangle$.
The probability of the transition from the initial
configuration to the final configuration is the probability $P(\leftarrow)$ to choose
a left move, times the probability $1-P_\leftarrow^\dagger(\tau)$ of no creation, times
the probability $P_\leftarrow^{kc}(a)$ to keep moving and create, times the probability $P_\leftarrow(\psi_R^\prime)$ to choose
the new state $\psi_R^\prime$, times the probability $1-P_\leftarrow^{kd}(\tau^\prime)$ to stop the move after
having destroyed, times the probability $P_\leftarrow^{L^\prime R^\prime}(\tau^\prime-\tau_R^\prime)$ to
shift the Green operator by $\tau^\prime-\tau_R^\prime$:
\begin{eqnarray}
  \nonumber S_{i\to f} &=&      P(\leftarrow)\big[1-P_\leftarrow^\dagger(\tau)\big]P_\leftarrow^{kc}(a)P_\leftarrow(\psi_R^\prime) \\
                      &\times& \big[1-P_\leftarrow^{kd}(\tau^\prime)\big]P_\leftarrow^{L^\prime R^\prime}(\tau^\prime-\tau_R^\prime)
\end{eqnarray}
The probability of the reverse move is exactly symmetric:
\begin{eqnarray}
  \nonumber S_{f\to i} &=&      P(\rightarrow^\prime)\big[1-P_\rightarrow^\dagger(\tau^\prime)\big]P_\rightarrow^{kc}(a)P_\rightarrow(\psi_L) \\
                      &\times& \big[1-P_\rightarrow^{kd}(\tau)\big]P_\rightarrow^{LR}(\tau_L-\tau)
\end{eqnarray}
The acceptance factor takes the form
\begin{eqnarray}
  \nonumber q_\leftarrow^{dc} & =    & \frac{\big[1-P_\rightarrow^{kd}(\tau)\big]\big(V_L-V_R\big)}{P(\leftarrow)\big[1-P_\leftarrow^\dagger(\tau)\big]\big[1-e^{-(\tau_L-\tau_R)(V_L-V_R)}\big]} \\
  \nonumber                   &\times& \frac{\big\langle\psi_L^a\big|\hat\mathcal G\hat\mathcal T\big|\psi_R^a\big\rangle P_\rightarrow^{kc}(a)}{\big\langle\psi_L^a\big|\hat\mathcal T\hat\mathcal G\big|\psi_R^a\big\rangle P_\leftarrow^{kc}(a)} \\
                              &\times& \frac{P(\rightarrow^\prime)\big[1-P_\rightarrow^\dagger(\tau^\prime)\big]\big[1-e^{-(\tau_L^\prime-\tau_R^\prime)(V_R^\prime-V_L^\prime)}\big]}{\big[1-P_\leftarrow^{kd}(\tau^\prime)\big]\big(V_R^\prime-V_L^\prime\big)},
\end{eqnarray}
and is written as a quantity that depends on the initial configuration, times
a quantity that depends on the intermediate configuration $a$, times a
quantity that depends on the final configuration. It is useful for the
remaining of the paper to define the intermediate acceptance factor,
\begin{equation}
  q_\leftarrow^{d-c}(a)=\frac{\big\langle\psi_L^a\big|\hat\mathcal G\hat\mathcal T\big|\psi_R^a\big\rangle P_\rightarrow^{kc}(a)}{\big\langle\psi_L^a\big|\hat\mathcal T\hat\mathcal G\big|\psi_R^a\big\rangle P_\leftarrow^{kc}(a)}.
\end{equation}

\subsubsection{Creation, destruction, creation}
We consider here the case where a left move is chosen, an operator is
created on the right of the Green operator, then the operator on its left
is destroyed, then a second operator is created on its right. Finally, a
time shift of the Green operator is performed. The sequence of configurations
is the following
\begin{enumerate}
  \item $P_i\propto\big\langle\psi_{L+1}\big|\hat\mathcal T(\tau_L)\big|\psi_L\big\rangle\big\langle\psi_L\big|\hat\mathcal G(\tau)\big|\psi_R\big\rangle$
  \item $\big\langle\psi_{L+1}^a\big|\hat\mathcal T(\tau_L^a)\big|\psi_L^a\big\rangle\big\langle\psi_L^a\big|\hat\mathcal G(\tau^a)\big|\psi_R^a\big\rangle\big\langle\psi_R^a\big|\hat\mathcal T(\tau_R^a)\big|\psi_{R-1}^a\big\rangle$
  \item $\big\langle\psi_L^b\big|\hat\mathcal G(\tau^b)\big|\psi_R^b\big\rangle\big\langle\psi_R^b\big|\hat\mathcal T(\tau_R^b)\big|\psi_{R-1}^b\big\rangle$
  \item $P_f\propto\big\langle\psi_L^\prime\big|\hat\mathcal G(\tau^\prime)\big|\psi_R^\prime\big\rangle\big\langle\psi_R^\prime\big|\hat\mathcal T(\tau_R^\prime)\big|\psi_{R-1}^\prime\big\rangle\big\langle\psi_{R-1}^\prime\big|\hat\mathcal T(\tau_{R-1}^\prime)\big|\psi_{R-2}^\prime\big\rangle$,
\end{enumerate}
Considering the intermediate configurations $a$ and $b$ between the intial
and final configurations, it is easy to show that
the corresponding acceptance factor can be written
\begin{equation}
  q_\leftarrow^{cdc}=q_\leftarrow^c \times q_\leftarrow^{c-d}(a) \times q_\leftarrow^{d-c}(b).
\end{equation}

\subsubsection{Destruction, creation, destruction}
We consider here the case where a left move is chosen, the operator on the
left of the Green operator is destroyed, then an operator is created on its
right. Finally a second operator on the left of Green operator is destroyed. The sequence of configurations
is the following
\begin{enumerate}
  \item $P_i\propto\big\langle\psi_{L+2}\big|\hat\mathcal T(\tau_{L+1})\big|\psi_{L+1}\big\rangle\big\langle\psi_{L+1}\big|\hat\mathcal T(\tau_L)\big|\psi_L\big\rangle\big\langle\psi_L\big|\hat\mathcal G(\tau)\big|\psi_R\big\rangle$
  \item $\big\langle\psi_{L+1}^a\big|\hat\mathcal T(\tau_L^a)\big|\psi_L^a\big\rangle\big\langle\psi_L^a\big|\hat\mathcal G(\tau^a)\big|\psi_R^a\big\rangle$
  \item $\big\langle\psi_{L+1}^b\big|\hat\mathcal T(\tau_L^b)\big|\psi_L^b\big\rangle\big\langle\psi_L^b\big|\hat\mathcal G(\tau^b)\big|\psi_R^b\big\rangle\big\langle\psi_R^b\big|\hat\mathcal T(\tau_R^b)\big|\psi_{R-1}^b\big\rangle$
  \item $P_f\propto\big\langle\psi_L^\prime\big|\hat\mathcal G(\tau^\prime)\big|\psi_R^\prime\big\rangle\big\langle\psi_R^\prime\big|\hat\mathcal T(\tau_R^\prime)\big|\psi_{R-1}^\prime\big\rangle$,
\end{enumerate}
Considering the intermediate configurations $a$ and $b$ between the intial
and final configurations, it is easy to show that
the corresponding acceptance factor can be written
\begin{equation}
  q_\leftarrow^{dcd}=q_\leftarrow^d \times q_\leftarrow^{d-c}(a) \times q_\leftarrow^{c-d}(b).
\end{equation}

\subsubsection{Generalization}
It is straighforward to show that the acceptance factors of the form
$q_\leftarrow^{cdcdc}$, $q_\leftarrow^{cdcdcdc}$, $q_\leftarrow^{cdcdcdcdc}\cdots$
(or  $q_\leftarrow^{dcdcd}$, $q_\leftarrow^{dcdcdcd}$, $q_\leftarrow^{dcdcdcdcd}\cdots$) can be expressed
as products of the acceptance factor $q_\leftarrow^c$ (or  $q_\leftarrow^d$) and the intermediate
factors $q_\leftarrow^{c-d}$ and $q_\leftarrow^{d-c}$.

In the same manner, the acceptance factors of the form
$q_\leftarrow^{cdcd}$, $q_\leftarrow^{cdcdcd}$, $q_\leftarrow^{cdcdcdcd}\cdots$
(or  $q_\leftarrow^{dcdc}$, $q_\leftarrow^{dcdcdc}$, $q_\leftarrow^{dcdcdcdc}\cdots$) can be expressed
as products of the acceptance factor $q_\leftarrow^{cd}$ (or  $q_\leftarrow^{dc}$) and the intermediate
factors $q_\leftarrow^{c-d}$ and $q_\leftarrow^{d-c}$.

\subsubsection{Simplification of the acceptance factors}
Here again it is possible to take advantage of the freedom that we have
for the choice of the probabilities $P(\leftarrow)$, $P_\leftarrow^\dagger$,
$P_\leftarrow^{kc}$, and $P_\leftarrow^{kd}$ (or  $P(\rightarrow)$, $P_\rightarrow^\dagger$, $P_\rightarrow^{kc}$, and $P_\rightarrow^{kd}$).
A proper choice of these probabilities can be done in order to allow us
to accept all moves, simplicity and generality being the leitmotiv of the SGF
algorithm.

For this purpose, we impose to all acceptance factors corresponding to
left (or  right) moves to be equal. This requires the intermediate
acceptance factors $q_\leftarrow^{c-d}$ and $q_\leftarrow^{d-c}$
(or  $q_\rightarrow^{c-d}$ and $q_\rightarrow^{d-c}$) to be equal to
1. This is realized if
\begin{eqnarray}
  && P_\leftarrow^{kc}=\alpha_c\min\bigg(1,\frac{\big\langle\hat\mathcal G\hat\mathcal T\big\rangle}{\big\langle\hat\mathcal T\hat\mathcal G\big\rangle}\bigg) \\
  && P_\rightarrow^{kc}=\alpha_c\min\bigg(1,\frac{\big\langle\hat\mathcal T\hat\mathcal G\big\rangle}{\big\langle\hat\mathcal G\hat\mathcal T\big\rangle}\bigg) \\
  && P_\leftarrow^{kd}=\alpha_d\min\big(1,e^{-\big(\tau_L^k-\tau_R^k\big)\big(V_R^k-V_L^k\big)}\big) \\
  && P_\rightarrow^{kd}=\alpha_d\min\big(1,e^{-\big(\tau_L^k-\tau_R^k\big)\big(V_L^k-V_R^k\big)}\big),
\end{eqnarray}
where $\alpha_c$ and $\alpha_d$ are optimization parameters belonging
to $\big[0;1[$. By tuning these parameters, the mean length
of the steps of the Green operator can be controlled. Note that we have explicitly excluded $1$ from the allowed
values for these optimization parameters. This is necessary for the Green operator to have a chance to end in a diagonal
configuration, $\big|\psi_L\big\rangle=\big|\psi_R\big\rangle$. Indeed, the choice $\alpha_c=\alpha_d=1$ would systematically lead to
values of $1$ for the probabilities $P^{kc}$ and $P^{kd}$ for diagonal configurations.
Therefore the Green operator would never stop in a diagonal configution, and no measurement could be done.
It is important here to note that the quantities  $\big\langle\hat\mathcal G\big\rangle$, $\big\langle\hat\mathcal G\hat\mathcal T\big\rangle$,
and $\big\langle\hat\mathcal T\hat\mathcal G\big\rangle$ are evaluated between the states on the left and the right
of the Green operator that are present at the moment where those quantities are needed, as well as for the
times indices $\tau_L^k$ and $\tau_R^k$ and the potentials $V_L^k$ and $V_R^k$.

All acceptance factors corresponding
to a given direction of propagation become equal if we choose for the
creation probabilities:
\begin{eqnarray}
  && P_\leftarrow^\dagger(\tau)=\frac{\big\langle\hat\mathcal G\hat\mathcal T\big\rangle}{\big\langle\hat\mathcal G\hat\mathcal T\big\rangle+\big\langle\hat\mathcal G\big\rangle\frac{\big[1-P_\rightarrow^{kd}\big](V_L-V_R)}{\big[1-P_\rightarrow^{kc}\big]\big[1-e^{-(\tau_L-\tau_R)(V_L-V_R)}\big]}} \\
  && P_\rightarrow^\dagger(\tau)=\frac{\big\langle\hat\mathcal T\hat\mathcal G\big\rangle}{\big\langle\hat\mathcal T\hat\mathcal G\big\rangle+\big\langle\hat\mathcal G\big\rangle\frac{\big[1-P_\leftarrow^{kd}\big](V_R-V_L)}{\big[1-P_\leftarrow^{kc}\big]\big[1-e^{-(\tau_L-\tau_R)(V_R-V_L)}\big]}},
\end{eqnarray}
Finally, all acceptances factors become independant of the direction of
propagation if we choose $P(\leftarrow)=\frac{r_\leftarrow(\tau)}{r_\leftarrow(\tau)+r_\rightarrow(\tau)}$
and $P(\rightarrow)=\frac{r_\rightarrow(\tau)}{r_\leftarrow(\tau)+r_\rightarrow(\tau)}$
with
\begin{eqnarray}
  r_\leftarrow(\tau)=\big[1-P_\rightarrow^{kc}\big]\frac{\big\langle\hat\mathcal G\hat\mathcal T\big\rangle}{\big\langle\hat\mathcal G\big\rangle}+\frac{\big[1-P_\rightarrow^{kd}\big](V_L-V_R)}{\big[1-e^{-(\tau_L-\tau_R)(V_L-V_R)}\big]} \\
  r_\rightarrow(\tau)=\big[1-P_\leftarrow^{kc}\big]\frac{\big\langle\hat\mathcal T\hat\mathcal G\big\rangle}{\big\langle\hat\mathcal G\big\rangle}+\frac{\big[1-P_\leftarrow^{kd}\big](V_R-V_L)}{\big[1-e^{-(\tau_L-\tau_R)(V_R-V_L)}\big]}.
\end{eqnarray}
As a result all moves can be accepted again, ensuring the maximum of simplicity of the algorithm.
We still have some freedom for the choice of the optimization parameters $\alpha_c$ and $\alpha_d$. This is discussed
in next section.

\section{Test and optimization of the algorithm}
From the central limit theorem, we know that the errorbar associated to any measured quantity
must decrease as the square root of the number of measurements, or equivalently, the square root
of the time of the simulation. Therefore it makes sense to define the efficiency $\mathcal E$ of a QMC algorithm
by
\begin{equation}
  \mathcal E(\Omega,\mathcal O)=\frac{1}{T(\Omega)\bigg(\Delta\mathcal O(\Omega)\bigg)^2},
\end{equation}
where $\Omega$ represents the set of all optimization parameters of the algorithm, $\mathcal O$ is the measured quantity
of interest, $T(\Omega)$ is the time of the simulation, and $\Delta\mathcal O(\Omega)$ is the errorbar associated to
the measured quantity $\mathcal O$. This definition ensures that $\mathcal E$
is independent of the time of the simulation. As a result, the larger $\mathcal E$ the more efficient the algorithm.

In the present case we have $\Omega=\big\lbrace\alpha_c,\alpha_d\big\rbrace$, while $\Omega=\emptyset$ for the original SGF algorithm.
It is useful here to realize that, by symmetry, the
mean values of $P_\leftarrow^{kc}$ and $P_\rightarrow^{kc}$ (and $P_\leftarrow^{kd}$ and $P_\rightarrow^{kd}$) must be equal.
Therefore we define $P^{kc}=\big\langle P_\leftarrow^{kc}\big\rangle=\big\langle P_\rightarrow^{kc}\big\rangle$ and
$P^{kd}=\big\langle P_\leftarrow^{kd}\big\rangle=\big\langle P_\rightarrow^{kd}\big\rangle$. It seems reasonable to impose
a condition of uniform sampling, $P^{kc}=P^{kd}$. This condition can be satisfied by adjusting dynamically the values of
$\alpha_c$ and $\alpha_d$ during the thermalization process. For this purpose we introduce a new optimization parameter $\alpha\in\big[0;1\big[$
and apply the following algorithm from time to time while thermalizing (we start with $\alpha_c=\alpha_d=\alpha$):
\begin{eqnarray}
  \nonumber && \textrm{Evaluate } P^{kc} \textrm{ and } P^{kd} \textrm{ over few iterations} \\
  \nonumber && \textrm{If } P^{kc}<P^{kd} \\
  \nonumber && \quad \textrm{then } \alpha_d\to\alpha_d\frac{P^{kc}}{P^{kd}} \\
  \nonumber && \quad \textrm{else } \alpha_c\to\alpha_c\frac{P^{kd}}{P^{kc}} \\
  \nonumber && \textrm{If } \alpha_c<\alpha_d \\
  \nonumber && \quad \textrm{then } \alpha_c=\frac{\alpha}{\alpha_d},\alpha_d=\alpha \\
  \nonumber && \quad \textrm{else } \alpha_d=\frac{\alpha}{\alpha_c},\alpha_c=\alpha
\end{eqnarray}
Thus we are left with the optimization parameter $\alpha$. In order to determine the optimal value, we have considered 2
different Hamiltonians $\hat\mathcal H_1$ and $\hat\mathcal H_2$, and evaluated the efficiency of the algorithm while
scanning $\alpha$. The first Hamiltonian we have considered describes free hardcore bosons and is exactly solvable,
\begin{equation}
  \hat\mathcal H_1=-t\sum_{\big\langle i,j\big\rangle}\bigg(a_i^\dagger a_j+a_j^\dagger a_i\bigg),
\end{equation}
where the sum runs over pairs of first neighboring sites and $t$ is the hopping parameter. The second Hamiltonian is highly non-trivial and describes a mixture
of atoms and diatomic molecules, with a special term allowing conversions between the two species \cite{RousseauFeshbach},
\begin{eqnarray}
  \nonumber \hat\mathcal H_2 &=& -t_a\sum_{\big\langle i,j\big\rangle}\bigg(a_i^\dagger a_j+a_j^\dagger a_i\bigg)-t_m\sum_{\big\langle i,j\big\rangle}\bigg(m_i^\dagger m_j+m_j^\dagger m_i\bigg) \\
  \nonumber                  &+& U_{aa}\sum_i \hat n_i^a\big(\hat n_i^a-1\big)+U_{mm}\sum_i \hat n_i^m\big(\hat n_i^m-1\big)+U_{am}\sum_i \hat n_i^a \hat n_i^m \\
  \label{TwoSpecies}         &+& D\sum_i \hat n_i^m+g\sum_i\bigg(m_i^\dagger a_i a_i+a_i^\dagger a_i^\dagger m_i\bigg),
\end{eqnarray}
where $a_i^\dagger$ and $a_i$ ($m_i^\dagger$ and $m_i$) are the creation and annihilation operators of atoms (molecules),
$t_a$, $t_m$, $U_{aa}$, $U_{mm}$, and $U_{am}$ are respectively the hopping parameter of atoms, the hopping parameter of molecules,
the atomic onsite interaction parameter, the molecular onsite interaction parameter, and the inter-species interaction parameter.
The conversion term is tunable via the parameter $g$ and does not conserve the number $N_a$ of atoms or the number $N_m$ of molecules.
However the total number of particles $N=N_a+2N_m$ is conserved and is the canonical constraint. The parameter $D$ allows
to control the ratio between the number of atoms and molecules.
The application of the SGF algorithm to the Hamiltonian (\ref{TwoSpecies}) is described in details in ref.\cite{RousseauSGF}. 
The changes coming with the directed update scheme are completely independent of the chosen Hamiltonian.

The following table shows the mean number of creations and destructions in one step, $\big\langle S(\alpha)\big\rangle$, and the relative efficiency $\mathcal E(\alpha,\mathcal O)/\mathcal E(\emptyset,\mathcal O)$ of the algorithm
applied to $\hat\mathcal H_1$ at half filling, for which we have measured the energy $E$, the superfluid density $\rho_s$, and the number of particles
in the zero momentum state $n(k=0)$:
\begin{table}
  \begin{tabular}{c c c c c}
    $\alpha$ & $\big\langle S(\alpha)\big\rangle$ & $\mathcal E(\alpha,E)$ & $\mathcal E(\alpha,\rho_s)$ & $\mathcal E(\alpha,n(0))$ \\
    \hline
    $0$          & 1.00 & 0.307400 & 0.487457  & 0.503105 \\
    $0.1$        & 1.10 & 0.774161 & 0.513633  & 0.805048 \\
    $0.5$        & 1.91 & 0.430843 & 3.771422  & 1.289757 \\
    $0.9$        & 7.00 & 0.977413 & 5.400997  & 6.629893 \\
    $0.95$       & 10.49 & 2.427874 & 10.688100 & 7.994883 \\
    $0.99$       & 17.49 & 1.286403 & 27.281408 & 1.327064 \\
    $0.9999$     & 20.93 & 0.818048 & 17.510068 & 1.059823 \\
    $0.999999$   & 21.00 & 0.710448 & 13.353809 & 0.779245 \\
    \label{Tab1}
  \end{tabular}
  \caption
    {
      Relative efficiency of the algorithm applied to $\hat\mathcal H_1$ at half filling for the energy, the superfluid
      density, and the number of particles in the zero momentum state.
    }
\end{table}

For $\hat\mathcal H_2$, we have used the parameters $t_a=1$, $t_m=1/2$, $U_{aa}=5$, $U_{mm}=5$,
$U_{am}=5$, $g=5$, $D=3$, and a density of particles $\rho=2$. The following tables shows $\big\langle S(\alpha)\big\rangle$,
and the relative efficiency of the algorithm for the energy $E$, the density of atoms and molecules $\rho_a$ and $\rho_m$,
the occupation of the zero momentum state for atoms and molecules $n_a(0)$ and $n_m(0)$, and the atomic
and molecular visibilities $\mathcal V_a$ and $\mathcal V_m$.
\begin{table}
  \begin{tabular}{c c c c c}
    $\alpha$ & $\big\langle S(\alpha)\big\rangle$ & $\mathcal E(\alpha,E)$ & $\mathcal E(\alpha,\rho_a)$ & $\mathcal E(\alpha,\rho_m)$ \\
    \hline
    $0$          & 1.00 & 1.086334 & 0.455569 & 1.670239 \\
    $0.1$        & 1.10 & 1.424308 & 0.506873 & 1.858339 \\
    $0.5$        & 1.88 & 2.813905 & 1.265620 & 4.640123 \\
    $0.9$        & 6.35 & 2.562529 & 5.999027 & 21.993900 \\
    $0.95$       & 8.99 & 2.335315 & 3.917233 & 14.361774 \\
    $0.99$       & 12.79 & 2.592328 & 1.721519 & 6.311612 \\
    \label{Tab2}
  \end{tabular}
  \caption
    {
      Relative efficiency of the algorithm applied to $\hat\mathcal H_2$ for the energy, and the density of atoms and molecules.
    }
\end{table}

\begin{table}
  \begin{tabular}{c c c c c}
    $\alpha$ & $\mathcal E(\alpha,n_a(0))$ & $\mathcal E(\alpha,n_m(0))$ & $\mathcal E(\alpha,\mathcal V_a)$ & $\mathcal E(\alpha,\mathcal V_m)$ \\
    \hline
    $0$          & 0.433382 & 0.234412 & 1.323720 & 0.239113 \\
    $0.1$        & 0.269700 & 0.181019 & 0.585183 & 0.248060 \\
    $0.5$        & 1.752466 & 2.806166 & 2.667114 & 1.357462 \\
    $0.9$        & 7.080124 & 5.638859 & 16.454676 & 4.482435 \\
    $0.95$       & 4.893878 & 3.757436 & 5.088775 & 2.248427 \\
    $0.99$       & 3.871723 & 2.341222 & 7.783268 & 1.279447 \\
    \label{Tab3}
  \end{tabular}
  \caption
    {
      Relative efficiency of the algorithm applied to $\hat\mathcal H_2$ for the occupation of the zero momentum state and the
      visibility of atoms and molecules.
    }
\end{table}

\begin{figure}[h]
  \centerline{\includegraphics[width=0.45\textwidth]{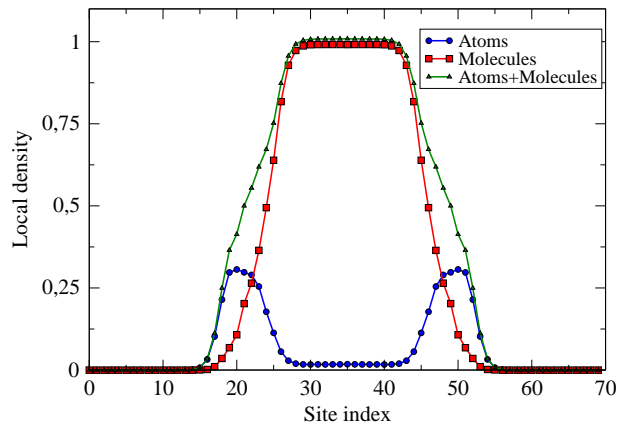}}
  \caption
    {
      (Color online) An example of density profiles obtained when adding the trapping potential (\ref{Trap}) to the Hamiltonian (\ref{TwoSpecies}).
      The errorbars are smaller than the symbol sizes, and are the biggest in the neighborhood of site indices 23 and 47
      where they equal the size of the symbols.
    }
  \label{Density}
\end{figure}

\begin{figure}[h]
  \centerline{\includegraphics[width=0.45\textwidth]{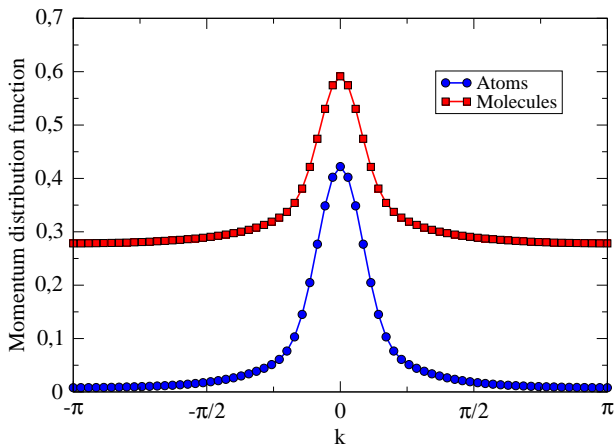}}
  \caption
    {
      (Color online) An example of momentum distribution functions obtained when adding the trapping potential (\ref{Trap}) to the Hamiltonian (\ref{TwoSpecies}).
      The errorbars are smaller than the symbol sizes, and are the biggest for $k=0$ where they equal the size of the symbols.
    }
  \label{Momentum}
\end{figure}

While the best value of $\alpha$ depends on the Hamiltonian which is considered and the measured quantity, it appears that
a good compromise is to choose $\alpha$ between $0.90$ and $0.99$. The improvment of the efficiency is remarkable.
In the following, we illustrate the applicability of the algorithm to problems with non-uniform potentials,
by adding a parabolic trap to the Hamiltonian (\ref{TwoSpecies}):
\begin{equation}
  \label{Trap} \hat\mathcal H_T=W_a\sum_i(i-L/2)^2 \hat n_i^a+W_m\sum_i(i-L/2)^2 \hat n_i^m
\end{equation}
The parameters $W_a$ and $W_m$ allow to control the curvature of the trap associated to atoms and molecules, respectively,
and $L$ is the number of lattice sites.
The inclusion of this term in the algorithm is trivial since only the values of the diagonal energies $V_L$ and $V_R$ are
changed. Figures (\ref{Density}) and (\ref{Momentum}) show the density profiles and momentum distribution functions
obtained for a system with $L=70$ lattice sites initially loaded with $50$ atoms and no molecules, and the parameters
$t_a=1$, $t_m=0.5$, $U_{aa}=4$, $U_{am}=12$, $U_{mm}=\infty$, $g=0.5$, $D=0$, $W_a=0.008$, $W_m=0.008$, and $\beta=20$.
The presented results have been obtained by performing $10^5$ updates for thermalization, and $2\times 10^5$ updates
with measurements (an update is to be understood as the occurence of a diagonal configuration). The time of the simulation
is about 8 hours on a cheap 32 bits laptop with 1GHz processor, with an implementation of the algorithm involving dynamical
structures with pointers (see ref.\cite{RousseauSGF}).

\section{Conclusion}
We have presented a directed update scheme for the SGF algorithm, which has the properties of keeping
the simplicity and generality of the original algorithm, and improves significantly its efficiency. 

\begin{acknowledgments}
I would like to express special thanks to Peter Denteneer for useful suggestions. This work is part of the research program of the "Stichting voor Fundamenteel Onderzoek der Materie (FOM),"
which is financially supported by the "Nederlandse Organisatie voor Wetenschappelijk Onderzoek (NWO)."
\end{acknowledgments}

\subsection{Appendix: Exponential random number generator}
We describe here how to generate numbers with the appropriate exponential distribution (\ref{ExponentialDistribution}).
Assuming that we have at our disposal a uniform random number generator that generates a random
variable $U$ with the distribution $\rho_U(u)=1$ for $u\in\big[0;1\big[$, we would like to find
a function $f$ such that the random variable $T=f(U)$ is generated with the distribution
\begin{equation}
  \rho_T^{\Delta\tau,\Delta V}(\tau)=\frac{\Delta V e^{-\tau\Delta V}}{1-e^{-\Delta\tau\Delta V}} \hspace{1cm} \tau\in\big[0;\Delta\tau\big[,
\end{equation}
where $\Delta\tau$ and $\Delta V$ are the parameters of the exponential distribution. Because of the
relation $T=f(U)$, the probability to find $T$ in the range $\big[\tau;\tau+d\tau\big[$ must be equal
to the probability to find $U$ in the range $\big[u;u+du\big[$. This implies the condition
\begin{equation}
  \rho_U(u)\big|du\big|=\rho_T^{\Delta\tau,\Delta V}(\tau)\big|d\tau\big|,
\end{equation}
with $\big|\frac{d\tau}{du}\big|=\pm\frac{df}{du}$. Thus we have
\begin{equation}
  \frac{\Delta V e^{-f(u)\Delta V}}{1-e^{-\Delta\tau\Delta V}}\frac{df}{du}=\pm 1.
\end{equation}
Taking the anti-derivative with respect to $u$ on both sides of the equation, we get
\begin{equation}
  \frac{-e^{-f(u)\Delta V}}{1-e^{-\Delta\tau\Delta V}}=\pm (u+C),
\end{equation}
where $C$ is a constant. This constant and the correct sign are determined by imposing the conditions
$f(0)=0$ and $f(1)=\Delta\tau$. As a result, if $u$ is a realization of $U$, then a realization of $T$
is given by
\begin{equation}
  \tau=-\frac{1}{\Delta V}\ln\big[1-u\big(1-e^{-\Delta\tau\Delta V}\big)\big].
\end{equation}


\begin{thebibliography}{10}
\bibitem{MonteCarlo} Nicholas Metropolis and S. Ulam,
Journal of the American statistical association, number 247, volume 44 (1949).
\bibitem{QMC1} D.C. Handscomb,
Proc. Cambridge Phil. Soc. 58, 594 (1962).
\bibitem{QMC2} M.H. Kalos,
Phys. Rev. 128, 1791 (1962).
\bibitem{QMC3} R. Blankenbecler, D.J. Scalapino and R.L. Sugar,
Phys. Rev. D 24, 2278 (1981).
\bibitem{Batrouni1992} G.G. Batrouni and R.T. Scalettar,
Phys. Rev. B {\bf 46}, 9051 (1992).
\bibitem{QMC4} W. von der Linden,
Phys. Rep. 220, 53 (1992).
\bibitem{QMC5} H.G. Evertz, G. Lana and M. Marcu,
Phys. Rev. Lett. 70, 875-879 (1993).
\bibitem{QMC6} D.M. Ceperley,
Rev. Mod. Phys. 67, 279 (1995).
\bibitem{QMC7} B.B. Beard and U.-J. Wiese,
Phys. Rev. Lett. 77 5130 (1996).
\bibitem{QMC8} ``Quantum Monte Carlo Methods in Physics and Chemistry'',
ed. M.P. Nightingale and C.J. Umrigar, NATO Science series C 525,
Kluwer Academic Publishers, Dordrecht, (1999).
\bibitem{Sandvik} A.W. Sandvik,
J. Phys. A {\bf 25}, 3667 (1992);
Phys. Rev. B {\bf 59}, 14157 (1999).
\bibitem{Prokofev} N.V. Prokof'ev, B.V. Svistunov, and I.S. Tupitsyn,
JETP Lett. {\bf 87}, 310 (1998).
\bibitem {Rigol03} M. Rigol, A. Muramatsu, G.G. Batrouni, and R.T. Scalettar,
Phys. Rev. Lett. {\bf 91}, 130403 (2003).
\bibitem{VanHoucke06} K. Van Houcke, S.M.A. Rombouts, and L. Pollet,
Phys. Rev. E {\bf 73},056703 (2006).
\bibitem {RousseauSGF} V.G. Rousseau,
Phys. Rev. E {\bf 77}, 056705 (2008).
\bibitem{Sandvik2002} A.W. Sandvik, S. Daul, R.R.P. Singh, and D.J. Scalapino2
Phys. Rev. Lett. {\bf 89}, 247201 (2002).
\bibitem{Rousseau2005} V.G. Rousseau, R.T. Scalettar, and G.G. Batrouni,
Phys. Rev. B {\bf 72}, 054524 (2005).
\bibitem{Metropolis} N. Metropolis, A.W. Rosenbluth, M.N. Metropolis, A.H. Teller, and E. Teller,
J. Chem. Phys. {\bf 21}, 1087 (1953).
\bibitem{Syljuasen} Olav F. Syljuasen, Anders W. Sandvik,
Phys. Rev. E {\bf 66}, 046701 (2002).
\bibitem{RousseauFeshbach} V.G. Rousseau and P.J.H. Denteneer,
Phys. Rev. A {\bf 77}, 013609 (2008).
\end{thebibliography}
\end{document}